\documentclass[prb,notitlepage,twocolumn]{revtex4}
\usepackage{amsmath}
\usepackage{amssymb}
\usepackage{bm}
\usepackage{graphicx}
\usepackage{times}

\begin{document}

\title{Nonlinear metasurfaces governed by bound states in the continuum}

\author{Kirill Koshelev$^{1,2}$}
\email{kirill.koshelev@anu.edu.au}
\author{Yutao Tang$^{3}$}
\author{Kingfai Li$^{3}$}
\author{Duk-Yong Choi$^{4,5}$}
\author{Guixin Li$^{3}$}
\author{Yuri Kivshar$^{1,2}$}

\affiliation{$^{1}$Nonlinear Physics Center, Australian National University, Canberra ACT 2601, Australia}
\affiliation{$^{2}$Department of Nanophotonics and Metamaterials, ITMO University, 197101 St.~Petersburg, Russian Federation}
\affiliation{$^{3}$Department of Material Science and Engineering, Shenzhen Institute for Quantum Science and Engineering, Southern University of Science and Technology, Shenzhen 518055, China}
\affiliation{$^{4}$Laser Physics Center, Australian National University, Canberra ACT 2601, Australia}
\affiliation{$^{5}$College of Information Science and Technology, Jinan University, Guangdong 510632, China}

\date{\today}

\begin{abstract}
Nonlinear nanostructured surfaces provide a paradigm shift in nonlinear optics with new ways to control and manipulate frequency conversion processes at the nanoscale, also offering novel opportunities for applications in photonics, chemistry, material science, and biosensing. Here, we develop a general approach to employ sharp resonances in metasurfaces originated from the physics of bound states in the continuum for both engineering and enhancing the nonlinear response. We study experimentally the third-harmonic generation from metasurfaces composed of symmetry-broken silicon meta-atoms and reveal that the harmonic generation intensity depends critically on the asymmetry parameter. We employ the concept of the critical coupling of light to the metasurface resonances to uncover the effect of radiative and nonradiative losses on the nonlinear conversion efficiency.
\end{abstract}

\maketitle

The study of nonlinear metasurfaces has emerged recently as a new exciting direction of research, and it attracted growing interest in the photonics community due to its ability to provide a paradigm shift in nonlinear optics~\cite{c1,c2}. Frequency conversion processes such as second- and third-harmonic generation are commonly realized in conventional nonlinear optics of macroscopic structures where high conversion efficiency is achieved by employing the concept of phase matching. Ultra-thin metasurfaces are capable to replace bulk elements yet providing reasonable conversion efficiencies~\cite{c2}.  Metasurfaces have the advantages over three-dimensional metamaterials in their lower optical losses and technical feasibility of fabrication, providing diverse functionalities such as efficient frequency conversion, fast optical switching, and modulation of light, and they can be employed for many applications in photonics, chemistry, material science, and biosensing. 

The first study of nonlinear effects with metasurfaces revealed inherently small conversion efficiencies in the plasmonic metamaterials based on split-ring resonators.  In such structures, nonlinear effects are driven by magnetic dipole resonances, but the first reported conversion efficiencies did not exceed $2\times10^{-11}$, for the second-harmonic generation, and $3\times10^{-12}$, for the third-harmonic generation~\cite{c3,c41,c42}.  The subsequent study of optically resonant dielectric metasurfaces driven by Mie resonances has improved substantially these numbers~\cite{c5,c6,c7,c8,c9,c10,c11} making nonlinear dielectric metasurfaces very attractive for practical applications.

\begin{figure*} [t]
\begin{minipage}{0.99\linewidth}
\begin{center}
\includegraphics[width=0.8\linewidth]{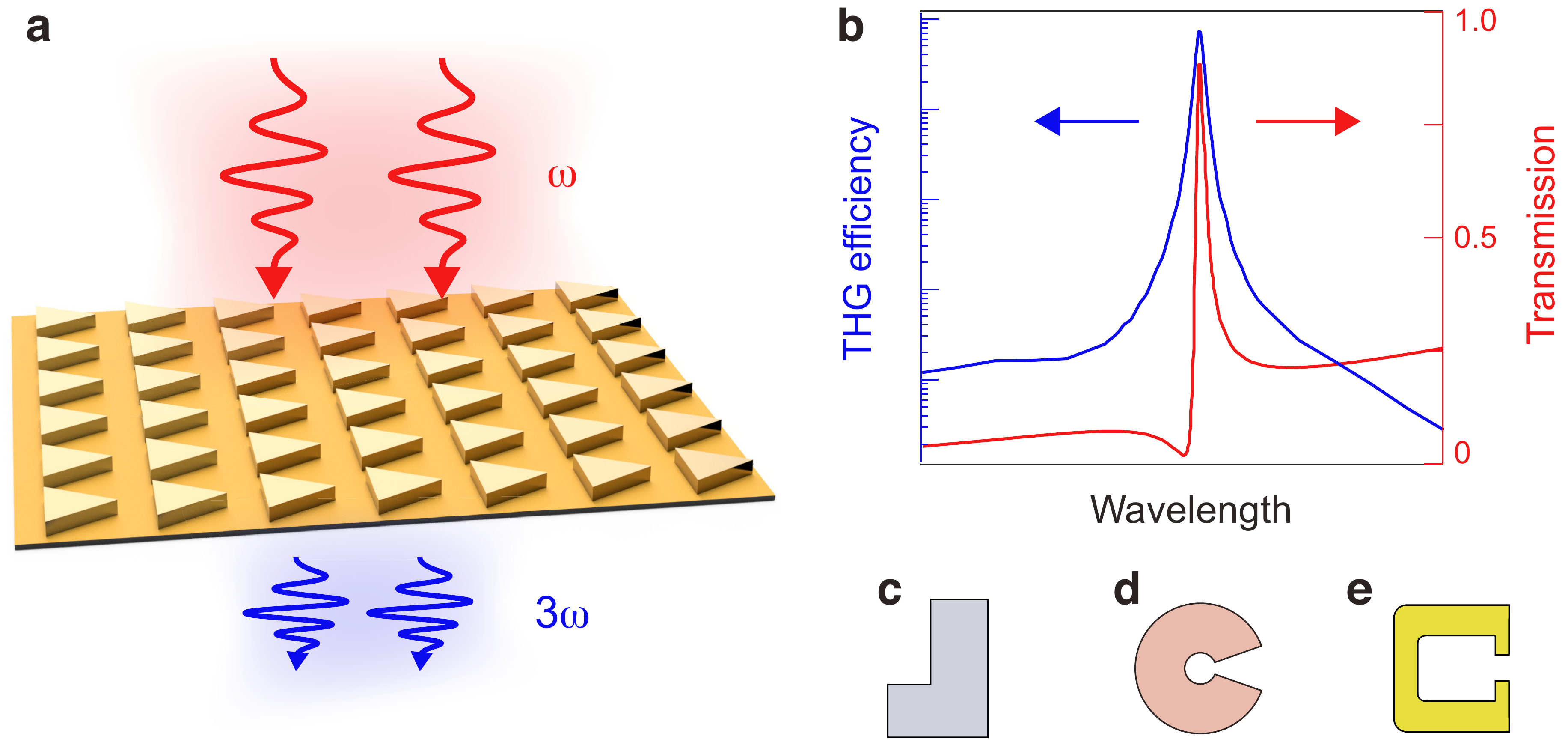}
\end{center}
\caption{{\bf Enhanced harmonic generation with asymmetric nonlinear metasurfaces.} (a) Schematic of the third-harmonic generation in a metasurface with asymmetric meta-atoms. (b) Schematic dependence of the conversion efficiency (blue) and transmission spectra (red) of a metasurface in the vicinity of a sharp quasi-BIC resonance. (c) Examples of the unit cells of asymmetric metasurfaces with broken-symmetry meta-atoms supporting high-Q resonances, as studied earlier in Refs.~\cite{c7,c8,c9}. } 
\label{fig:1}
\end{minipage}
\end{figure*}

All-dielectric resonant meta-optics has emerged recently as a novel research field driven by its exceptional applications for creating low-loss nanoscale metadevices~\cite{c12}. The tight confinement of the local electromagnetic fields and multiple interferences available in resonant high-index dielectric nanostructures and metasurfaces can boost many optical effects and offer novel opportunities for the subwavelength control of light-matter interactions. In particular, recently emerged concept of bound states in the continuum (BICs) in nanophotonics enables a simple approach to achieve high-Q resonances (or quasi-BICs) for various platforms ranging from individual dielectric nanoparticles~\cite{c13} to periodic arrangements of subwavelength resonators such as metasurfaces or chains of particles~\cite{c14,c15,c16}. Moreover, very recently it was revealed~\cite{c17} that metasurfaces created by seemingly different lattices of dielectric meta-atoms with broken in-plane inversion symmetry can support sharp high-Q resonances arising from a distortion of symmetry-protected BICs. For applications in nonlinear optics, with similarity to the recently studied isolated dielectric resonators~\cite{c18}, we wonder if such high-Q resonances can boost substantially nonlinear parametric processes in dielectric metasurfaces such as third-harmonic generation (THG), as shown schematically in Figs. 1(a, b).  Indeed, we notice that some earlier studies suggest that the use of certain types of meta-atoms with broken symmetry, such as those shown in Fig.~\ref{fig:1}d~\cite{c9}, Fig.~\ref{fig:1}d~\cite{c10} and Fig.~\ref{fig:1}e~\cite{c7} can lead to enhanced nonlinear effects in metasurfaces, often associated with Fano resonances. A rigorous theory of such asymmetric periodic structures developed in Ref.~\cite{c17} revealed a close link between the Fano resonances and BICs, so we expect that the BIC-assisted resonances may provide a generic enhancement for nonlinear metasurface, including those discussed in Ref.~\cite{c17}.

In this Letter, we suggest and develop a novel concept for enhancing and tailoring nonlinear response of dielectric metasurfaces with broken in-plane symmetry exploiting the powerful physics of {\it bound states in the continuum}. We apply our approach to silicon-based metasurfaces generating third-harmonic fields. We fabricate a set of metasurfaces varying the parameters of meta-atoms and analyse how the THG efficiency depends on the degree of asymmetry. We demonstrate that the interplay of radiative and nonradiative losses can control the intensity of the third-harmonic response. In particular, we reveal that tuning the metasurface parameters to the regime of the critical coupling, when the contributions of two loss mechanisms coincide, allows to achieve the maximum efficiency of the frequency conversion in metasurfaces. 

\begin{figure*} [t]
\begin{minipage}{0.99\linewidth}
\begin{center}
\includegraphics[width=0.85\linewidth]{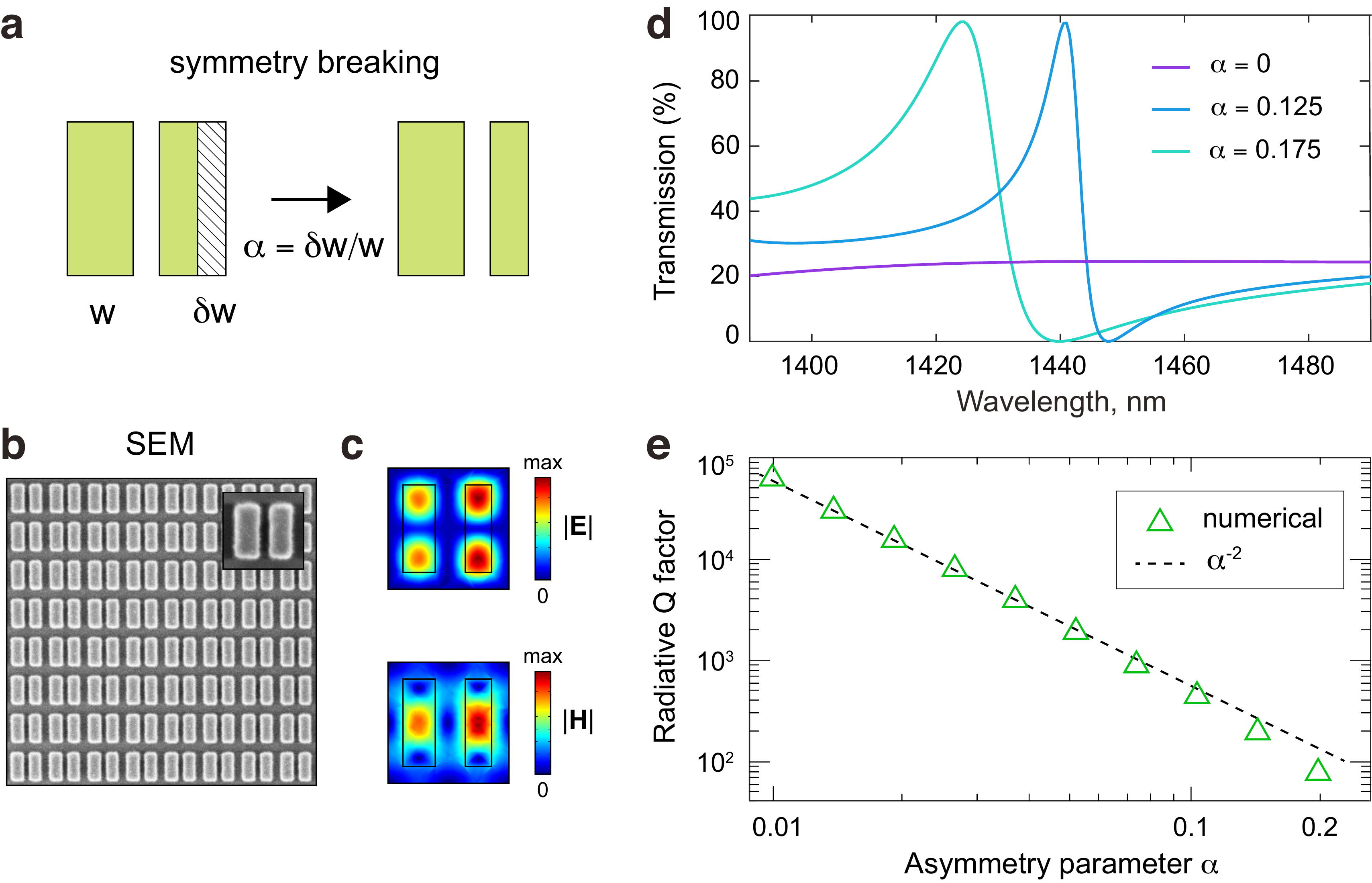}
\end{center}
\caption{{\bf High-Q resonances in asymmetric metasurfaces associated with quasi-BIC modes.} (a) Design of the unit cell and the definition of the asymmetry parameter. (b) Top-view SEM image of the silicon metasurface placed on a glass substrate. The inset shows the high-resolution SEM image of the meta-atom. (c) Near-field distributions of the electric and magnetic fields of the quasi-BIC mode. (d) Evolution of the transmission spectra with respect to the metasurface asymmetry. (e) Dependence of the radiative Q factor on the asymmetry parameter. The dashed line shows an inverse quadratic fitting.} 
\label{fig:2}
\end{minipage}
\end{figure*}

We consider a meta-atom of the designed metasurface in the form of an asymmetric pair of bars, as illustrated in Fig.~\ref{fig:2}a; the rectangular bars have the widths $w$ and $w-\delta w$, respectively. The asymmetry of the unit cell is controlled by the difference in bar widths, and it is characterized by the asymmetry parameter $\alpha=\delta w/w$.  We perform numerical analysis of linear response and eigenmode spectra of the asymmetric metasurface using a commercially available software for full-wave electromagnetic simulations (see Methods below). We consider the metasurface as an infinite structure with perfect periodicity and all geometrical and material parameters taken from the experimental data below. A symmetric metasurface with $\alpha=0$ supports a BIC at $1465$ nm with infinite quality factor (the Q factor) protected by the in-plane symmetry of the unit cell~\cite{c19}. The BIC is not manifested in the transmission spectrum shown in Fig.~\ref{fig:2}d at $\alpha=0$ due to the symmetry incompatibility with the modes of free space. For broken-symmetry metasurfaces, the transmission curves reveal a sharp resonance with a Fano lineshape associated with a quasi-BIC with high Q factor. Figure~\ref{fig:2}c demonstrates the near-field distribution for quasi-BIC resembling the interaction of two oppositely directed magnetic dipoles with slightly dissimilar amplitudes. The evolution of the quasi-BIC radiative Q factor $Q_{\rm r}$ on the asymmetry parameter of the meta-atom follows the inverse quadratic law for small $\alpha\le0$~\cite{c17}, as shown in Fig.~\ref{fig:2}e,
\begin{equation}
    Q_{\rm r}=Q_0\alpha^{-2},
\label{eq:1}
\end{equation}
where $Q_0$ is a constant determined by the metasurface design being independent on $\alpha$. For larger $\alpha$ the decrease of $Q_{\rm r}$ is going faster because the deviation from the symmetric unit cell cannot be considered as a weak perturbation.

To demonstrate how the meta-atom asymmetry shapes the TH response, we fabricate a set of six silicon metasurfaces on a glass substrate with different asymmetry parameters using the electron beam lithography with the details provided in Methods. Figure~\ref{fig:2}b shows the scanning electron microscope (SEM) image of the fabricated pattern, and the inset shows high-resolution SEM image for a single meta-atom. The pattern size of each fabricated metasurface is $100$ $\mu$m $\times$ $100$ $\mu$m, the period $655$ nm, height is $538$ nm, bar length is $500$ nm, larger bar width is $200$ nm, and the distance between bar centres is $320$ nm. The width of the smaller bar varies in the range $155$, $165$, $170$, $180$, $185$, $195$ nm, which corresponds to the asymmetry parameter of   $\alpha=0.225$, $0.175$, $0.15$, $0.1$, $0.075$, and $0.05$, respectively. We pump the metasurfaces from air in the wavelength range from $1370$ to $1470$ nm with femtosecond pulses from a coherent optical parametric amplifier (see Methods). The generated TH signal is collected in the forward direction by a high numerical aperture objective.

\begin{figure*} [t]
\begin{minipage}{0.99\linewidth}
\begin{center}
\includegraphics[width=0.95\linewidth]{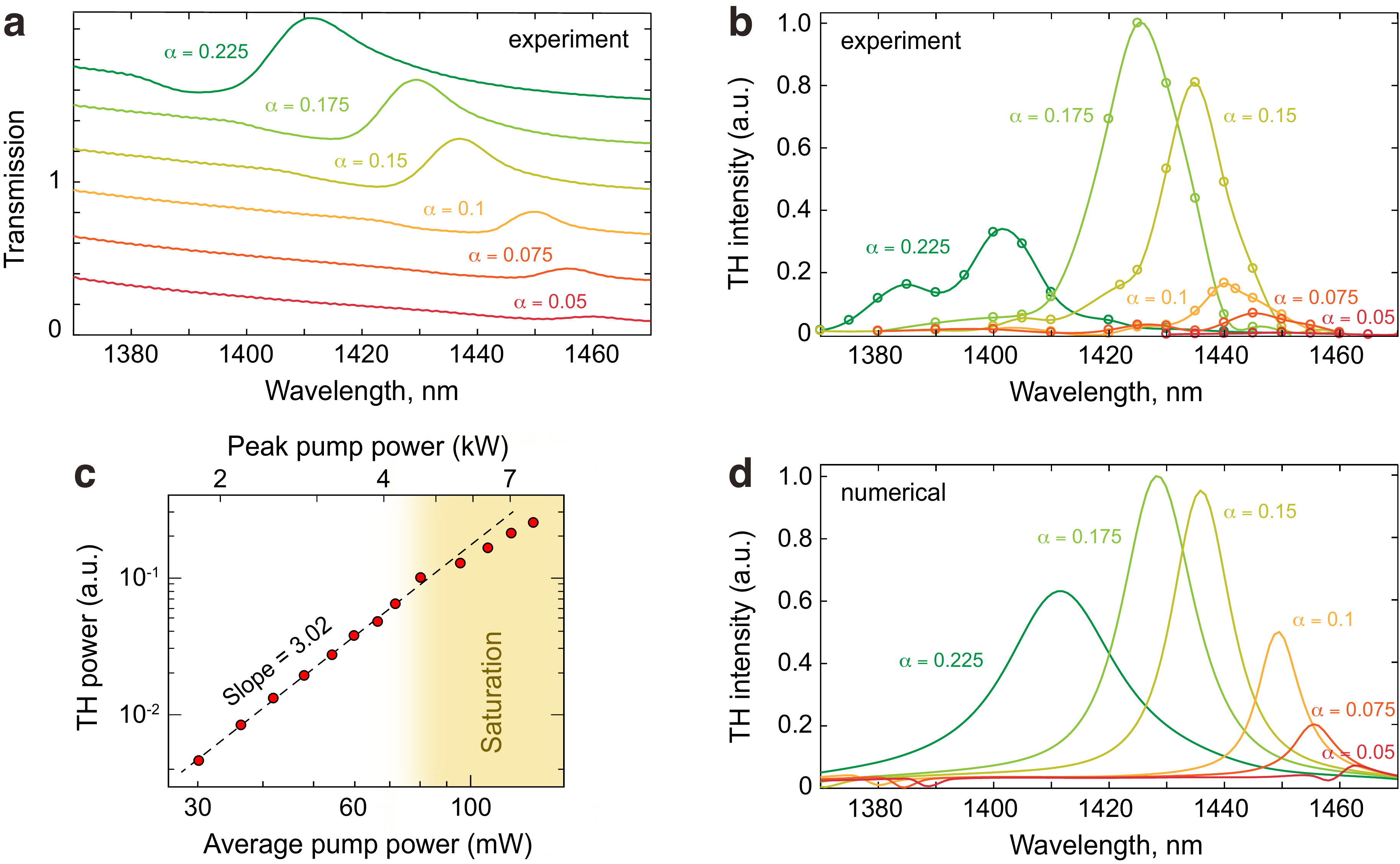}
\end{center}
\caption{{\bf Experimental results}. (a) Evolution of the measured transmission spectra vs. meta-atom asymmetry $\alpha$. Spectra are shifted relatively by $0.3$ units. (b) Observation of THG in asymmetric metasurfaces in the regime of the quasi-BIC resonance. The open circles show the experimental data, the solid curves provide a guide for eyes. (c) Dependence of the output TH power on the average and peak input power for the sample with $\alpha=0.15$. The red dots show the measured data, the black dashed line is a fit to the data with a third-order power function. A yellow area covers the range of input powers for which the TH power dependence deviates from a cubic law. (d) Numerically simulated dependence of the TH intensity with respect to the asymmetry parameter $\alpha$.
} 
\label{fig:3}
\end{minipage}
\end{figure*}

Figure~\ref{fig:3} illustrates our experimental results for the THG enhancement in the broken-symmetry metasurfaces driven by BICs. The measured transmission spectra demonstrate a quasi-BIC resonance with a Fano lineshape which resonant wavelength and linewidth strongly depend on $\alpha$. The mode positions in the experimental linear spectra agree well with the simulation (see Fig.~\ref{fig:2}d), though the Q factor and peak transmittance are reduced, which is expected due to non-radiative losses in the fabricated sample induced by surface roughness, structural disorder, and a finite size of the sample~\cite{c20,c21}. We estimate the value of the nonradiative Q factor as $Q_{\rm nr}=175$ by using the fitting procedure (see Methods). Figure~\ref{fig:3}b shows the evolution of the measured TH intensity with respect to the meta-atom asymmetry. The highest TH intensity is observed for the intermediate value of the asymmetry parameter $\alpha=0.175$, whereas for larger and smaller  the nonlinear signal is weaker. The dependence of the output power on the pump power for the metasurface with $\alpha=0.15$ is shown in Fig.~\ref{fig:3}c. For average input powers below $80$ mW, the TH power clearly follows the third-order power law, while for higher pump intensities the dependence shows a typical saturation behaviour. 

To simulate the nonlinear response of the fabricated metasurfaces, we introduce artificial nonradiative losses characterized by $Q_{\rm nr}=175$ to the numerical model (see Methods). The simulated TH spectra shown in Fig. 3d demonstrate good agreement with the experimental data. Considering the simulated directivity pattern of the TH wave, we conclude that only $5\%$ of the total TH intensity is emitted to the zero-diffraction order in the forward direction, while the rest of signal is generated in the backward direction and scattered in high-order diffraction channels. At the same time, the effective area of illumination is about $40\%$ which gives an estimate for the amount of the pump power coupled to the metasurface. Using these assumptions, we calculate the intrinsic experimental efficiency of THG for the sample with $\alpha=0.175$ demonstrating the best performance. The estimated overall conversion efficiency $P_{3\omega}/P_{\omega}$ is $10^{-6}$ for the average pump power of $P_{\omega}=130$ mW, which is similar to the previous reports on TH efficiency in Si nanostructures in the vicinity of dipole modes, composite resonances or anapoles~\cite{c5,c7,c11,c18,c22,c23,c24}. The maximum value of TH signal is affected dramatically by the non-radiative losses due to structural imperfections, disorder and finite size of the fabricated metasurface, which can be significantly decreased by enlarging the sample footprint~\cite{c25} and improving the fabrication quality.

The question about the interplay between radiative and nonradiative losses attracts a wide attention in various areas of optics, especially, for describing hybrid resonator-coupler systems, where the condition of equality of internal resonator losses and coupling losses, called critical coupling, is a fundamental feature~\cite{c26,c27,c28}. The fulfillment of this condition is favorable since it allows for complete exchange of energy between a propagating mode in a coupler device and the given resonator mode. The critical coupling principle is studied in application to plasmonics, metamaterials, and other areas~\cite{c29}, and here we employ it for describing nonlinear response of subwavelength metasurfaces governed by geometric and material resonances. We develop a model based on the temporal coupled mode theory (TCMT)~\cite{c30} and use it for describing the intensity of the third-order nonlinear signal.

The fabricated broken-symmetry metasurfaces support a quasi-BIC mode lying in the wavelength range from $1400$ to $1460$ nm (see Fig.~\ref{fig:3}a) characterized by the radiative and nonradiative Q factors, where $Q_{\rm r}$ depends on the asymmetry of the meta-atom according to Eq.~(\ref{eq:1}). The total Q factor $1 / Q=1 / Q_{\rm r}+1 / Q_{\rm nr}$ is also determined by the asymmetry
\begin{equation}
Q(\alpha)=\frac{Q_{\rm nr}}{\alpha^{2} / \alpha_{\rm cr}^{2}+1}
\label{eq:2}
\end{equation}
Here, $\alpha_{\rm cr}$ is the critical value of the asymmetry parameter
\begin{equation}
\alpha_{\rm cr}=\left(\frac{Q_{0}}{Q_{\rm nr}}\right)^{1 / 2}
\label{eq:3}
\end{equation}

The TCMT predicts that the amplitude $A_\omega$ of the quasi-BIC mode at the resonant wavelength is determined by the Q factor and the pump intensity $P_\omega$~\cite{c31}
\begin{equation}
\left|A_{\omega}\right|^{2} \propto[\alpha Q(\alpha)]^{2} P_{\omega}
\label{eq:4}
\end{equation}
The maximum amplitude is achieved exactly at the critical coupling condition $Q_{\rm r}=Q_{\rm nr}$, when $\alpha=\alpha_{\rm cr}$.  The intensity of nonlinear TH signal $P_{3\omega}$ is proportional to $\left|A_{\omega}\right|^{6}$ which at the pump resonance gives
\begin{equation}
P_{3 \omega} \propto[\alpha Q(\alpha)]^{6} P_{\omega}^{3}
\label{eq:5}
\end{equation}
Equation~(\ref{eq:5}) shows that the highest THG conversion efficiency $P_{3\omega}/P_{\omega}$ is achieved in the critical coupling regime, and the maximum value of efficiency scales with respect to the magnitude of nonradiative losses as $Q_{\rm nr}^3$.

\begin{figure*} [t]
\begin{minipage}{0.99\linewidth}
\begin{center}
\includegraphics[width=0.9\linewidth]{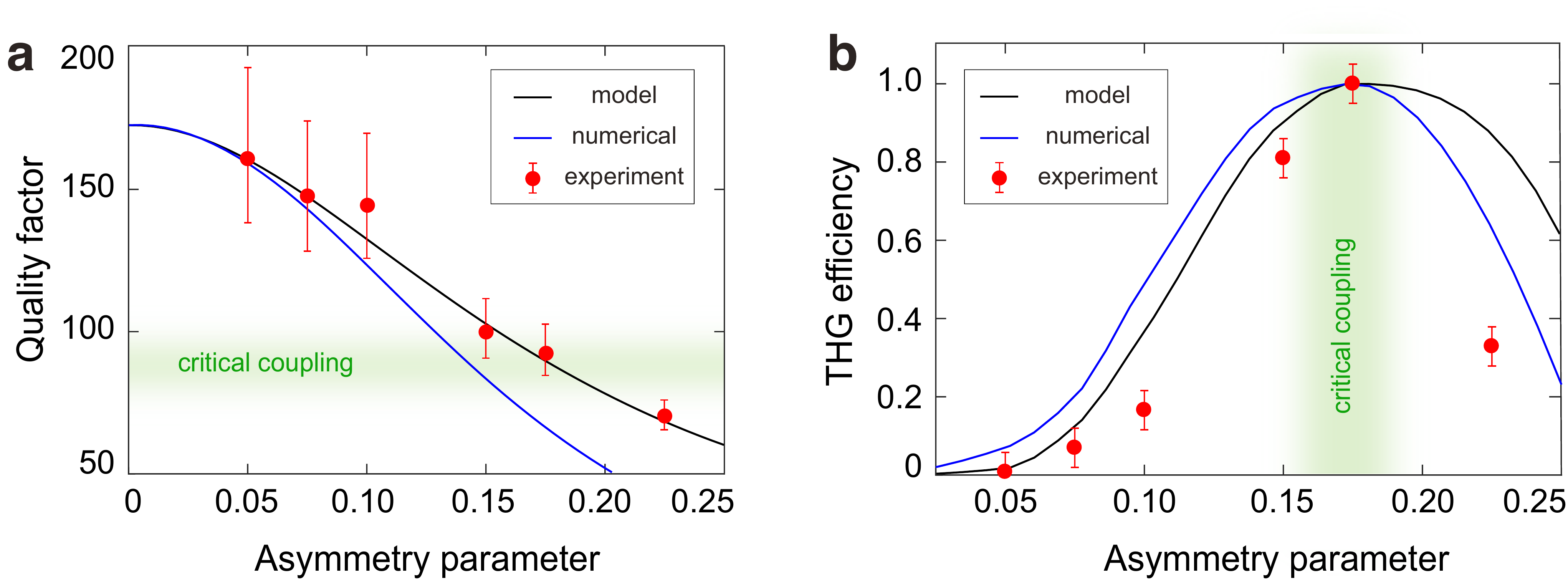}
\end{center}
\caption{{\bf Effect of nonradiative losses and critical coupling: Theory vs. experiment}.  Dependence of (a) the total Q factor and (b) THG conversion efficiency on the meta-atom asymmetry parameter $\alpha$ based on the experimental data (red marks), numerical calculations (blue solid) and the analytical model (black solid). The error bars show the magnitude of the error due to the fitting procedure. The analytical model is based on Eq.~(\ref{eq:2}) for the plot (a) and Eq.~(\ref{eq:5}) for the plot (b). The range of parameters satisfying the critical coupling condition is shown by a green blurred area.
} 
\label{fig:4}
\end{minipage}
\end{figure*}

A comparison of experimental, numerical, and analytical results for the dependence of the Q factor on the asymmetry parameter is illustrated in Fig.~\ref{fig:4}a. The total Q factor is extracted from the experimental data by the fitting procedure, with an accuracy shown with the error bars. The value of nonradiative Q factor $Q_{\rm nr}=175$ is extrapolated as $Q_{\rm nr}=Q(0)$. The numerical dependence $Q(\alpha)$ is calculated by $Q_{\rm nr}$ (see Methods). Using the eigenmode analysis for small $\alpha$ and Eq.~(\ref{eq:1}), we calculate the coefficient $Q_0=5.67$, which gives the analytical estimate $\alpha_{\rm cr}=0.18$. The analytical dependence based on Eq.~(\ref{eq:2}) with given parameters $Q_{\rm nr}$ and $\alpha_{\rm cr}$ is shown with a black solid line. The green area illustrates the critical coupling condition $Q_{\rm r}=Q_{\rm nr}$ corresponding to the range of asymmetry parameters $0.15\le\alpha_{\rm cr}\le0.2$, which is blurred due to an error in calculating $Q_{\rm nr}$. The critical coupling regime is achieved for the fabricated metasurface with $\alpha=0.175$ which agrees well with the analytical estimate for $\alpha_{\rm cr}$.

Figure~\ref{fig:4}b demonstrates the evolution of the peak values of the THG conversion efficiency for different asymmetries based on the experimental data, numerical calculations, and our analytical model. The experimental and numerical curves illustrate the values from Figs. 2(b, d), respectively, the analytical dependence is based on Eq.~(\ref{eq:5}), and all data are normalized independently. The sample with $\alpha=0.175$ demonstrates the best performance, as is predicted by the critical coupling model. The obtained results show high sensitivity of the nonlinear signal on the meta-atom asymmetry:  $20$ nm change of $\delta w$ results in a decrease of the THG efficiency by at least one order of magnitude.

In summary, we have developed a new approach for engineering the nonlinear response of dielectric metasurfaces composed of meta-atoms with broken in-plane symmetry closely associated with the physics of bound states in the continuum. We have employed our approach for silicon-based metasurfaces to tailor the THG efficiency depending on the degree of the unit cell asymmetry. We have demonstrated that performance of nonlinear metasurfaces can be improved by exploiting the interplay of radiative and nonradiative losses near the critical coupling condition.  The similar approach can be applied to the case of the second-harmonic generation from broken-symmetry III-V semiconductor metasurfaces~\cite{c9} or nonlinear metasurfaces composed of arrays of zinc oxide nanoresonators designed for the nonlinear optical generation of VUV light~\cite{c32}. Thus, we believe that our general approach paves the way to smart engineering of sharp resonances in metasurfaces with inherent absorption losses and fabrication imperfections for nonlinear meta-optics and nanophotonics, towards efficient frequency conversion, fast optical switching, and modulation of light.

\begin{acknowledgments}
The authors thank Sergey Kruk and Andrey Bogdanov for useful discussions and suggestions. G.L. was supported by the Guangdong Provincial Innovation and Entrepreneurship Project (2017ZT07C071), Applied Science and Technology Project of Guangdong Science and Technology Department (2017B090918001), and National Natural Science Foundation of China (11774145). Y.K. acknowledges a financial support from the Australian Research Council and the Strategic Fund of the Australian National University. K.K. was supported by the Russian Science Foundation (17-12-01581) and the Foundation for the Advancement of Theoretical Physics and Mathematics BASIS.
\end{acknowledgments}

\appendix

\section*{METHODS}

\subsection*{Sample fabrication.} For fabrication of metasurfaces composed of silicon bar pairs on the glass substrate we first deposit thin-films of hydrogenated amorphous silicon (a-Si:H) with a thickness of $538$ nm, using plasma-enhanced chemical vapor deposition (PECVD) at a temperature of $250$ $^{\rm o}$C on standard microscope coverslips. Next, the substrates are spin-coated with the positive-tone electron-beam resist (ZEP520A from Zeon Chemicals). The pattern of bar pairs is then defined by electron-beam lithography (EBL). A $50$-nm thick Al layer was subsequently deposited by e-beam evaporation (Temescal BJD-2000), accompanied by a lift-off process in which the sample is soaked in a resist remover (ZDMAC from ZEON Co.).  An array of remaining rectangular Al patterns was used as the etch mask to transfer the designed pattern into the a-Si:H film through inductively coupled plasma-reactive ion etching (Plasmalab System 100, Oxford). As etch gases, we used SF6 ($2.5$~sccm) and CHF3 ($50$~sccm). Etching was performed at $20$~$^{\rm o}$C with $15$~mTorr at $500$~W induction power and $15$ W bias power. Finally, residual Al was removed using Al wet etchant (H3PO4:HNO3:CH3COOH:H2O).

\subsection*{Optical experiments.}  For linear the optical measurement, a halogen white light is focused onto to the silicon metasurface after passing through a Glan-Thompson linear polarizer. The transmitted light with both co-polarization and cross-polarization is then spectrally resolved by using NIR Andor spectrometer (KYMERA-328i). In the nonlinear optical measurements, we use a spectrally tunable fs laser source (repetition frequency: $80$ MHz, pulse duration: $200$ fs). The considered spectral range of the pump laser is from $1370$ to $1470$ nm. The linearly polarized fundamental wave with a spot size of $20$ $\mu$m in diameter was normally incident onto the silicon metasurface after passing through an objective lens (NA = $0.1$). The THG signals in transmission direction are collected by an infinity-corrected objective lens (NA= $0.25$) onto an Andor spectrometer (SP500i) with a EMCCD detector. The origin of the TH signal is verified by the direct measurement of its spectrum and its power dependence which is in a cubic manner.

\subsection*{Numerical calculations.}  For numerical simulations of the linear and nonlinear response, we used the finite-element-method solver in COMSOL Multiphysics in the frequency domain. For simulations of the eigenmode spectra, we used the eigenmode solver in COMSOL Multiphysics. All calculations were realized for a metasurface on a semi-infinite substrate surrounded by a perfect matched layer mimicking an infinite region. The simulation area is the unit cell extended to an infinite metasurface by using the Bloch boundary conditions. All material properties including absorption losses are extracted from the ellipsometry data. The incident field is a plane wave in the normal excitation geometry polarized along the long side of the bars. For linear calculations the non-radiative losses were limited to the actual absorption losses. For nonlinear calculations, the additional non-radiative losses corresponding to $Q_{\rm nr}=175$ were artificially introduced via correction to the extinction coefficient $\Delta k=n/(2Q_{\rm nr})$, where $n$ is the refractive index.  The nonlinear simulations of TH response we employed within the undepleted pump approximation, using two steps to calculate the intensity of the radiated nonlinear signal. First, we simulated the linear scattering at the pump wavelength, and then obtained the nonlinear polarization induced inside the meta-atom. Then, we employed the polarization as a source for the electromagnetic simulation at the harmonic wavelength to obtain the generated TH field. The nonlinear susceptibility function $\chi^{(3)}$ was considered as a tensor corresponding to the cubic crystallographic point group with $\chi^{(3)} = 2.45\times10^{-19}$~m$^2$/V$^2$. The extraction of Q factors from the experimental transmission spectra was based on the single-peak fitting to a Fano lineshape using the Levenberg-Marquardt algorithm.

\bibliography{THG_BIC}

\end{document}